\documentclass[twocolumn,tighten,resetfootnote]{aastex7}
\usepackage{hyperref}
\usepackage{apjfonts}
\usepackage{apjfonts} 
\usepackage{graphicx} 
\usepackage{amssymb,amsmath} 
\usepackage{color} 

\renewcommand{\dbltextfloatsep}{4mm}
\renewcommand{\textfloatsep}{4mm}

\newcommand{\chandra}{\textit{Chandra}}

\newcommand{\xmm}{\textit{XMM-Newton}}

\newcommand{\xrism}{\textit{XRISM}}
\newcommand{\resolve}{{Resolve}}
\newcommand{\hitomi}{\textit{Hitomi}}

\newcommand{\am}{$^{\prime}$}
\renewcommand{\deg}{$^{\circ}$}

\newcommand{\kms}{~km~s$^{-1}$}
\def\kmsmpc{km$\;$s$^{-1}\,$Mpc$^{-1}$}
\def\cmsq{cm$^{-2}$}
\def\lax{\lesssim}
\def\gax{\gtrsim}

\begin{document}

\vspace*{-8mm}\hfill{\footnotesize ApJ Letters in press; submitted 2025 April 2; accepted 2025 April 29}
\vspace*{8mm}

\righthead{\xrism\ velocities for the Coma cluster}
\lefthead{\xrism\ Collaboration}

\title{\Large \xrism\ forecast for the Coma cluster: stormy, with a steep 
power spectrum}

\suppressAffiliations
\correspondingauthor{Maxim Markevitch (maxim.markevitch@nasa.gov).\\
\hspace*{-4mm}Authors' affilliations are given at the end of the preprint.}

\collaboration{0}{XRISM Collaboration:}

\author[0000-0003-4721-034X]{Marc Audard}
\affiliation{Department of Astronomy, University of Geneva, Versoix CH-1290, Switzerland} 
\email{Marc.Audard@unige.ch}

\author{Hisamitsu Awaki}
\affiliation{Department of Physics, Ehime University, Ehime 790-8577, Japan}
\email{awaki@astro.phys.sci.ehime-u.ac.jp}

\author[0000-0002-1118-8470]{Ralf Ballhausen}
\affiliation{Department of Astronomy, University of Maryland, College Park, MD 20742, USA}
\affiliation{NASA / Goddard Space Flight Center, Greenbelt, MD 20771, USA}
\affiliation{Center for Research and Exploration in Space Science and Technology, NASA / GSFC (CRESST II), Greenbelt, MD 20771, USA}
\email{ballhaus@umd.edu}

\author[0000-0003-0890-4920]{Aya Bamba}
\affiliation{Department of Physics, University of Tokyo, Tokyo 113-0033, Japan}
\email{bamba@phys.s.u-tokyo.ac.jp}

\author[0000-0001-9735-4873]{Ehud Behar}
\affiliation{Department of Physics, Technion, Technion City, Haifa 3200003, Israel}
\email{behar@physics.technion.ac.il}

\author[0000-0003-2704-599X]{Rozenn Boissay-Malaquin}
\affiliation{Center for Space Sciences and Technology, University of Maryland, Baltimore County (UMBC), Baltimore, MD, 21250 USA}
\affiliation{NASA / Goddard Space Flight Center, Greenbelt, MD 20771, USA}
\affiliation{Center for Research and Exploration in Space Science and Technology, NASA / GSFC (CRESST II), Greenbelt, MD 20771, USA}
\email{rozennbm@umbc.edu}

\author[0000-0003-2663-1954]{Laura Brenneman}
\affiliation{Center for Astrophysics | Harvard-Smithsonian, Cambridge, MA 02138, USA}
\email{lbrenneman@cfa.harvard.edu}

\author[0000-0001-6338-9445]{Gregory V.\ Brown}
\affiliation{Lawrence Livermore National Laboratory, Livermore, CA 94550, USA}
\email{brown86@llnl.gov}

\author[0000-0002-5466-3817]{Lia Corrales}
\affiliation{Department of Astronomy, University of Michigan, Ann Arbor, MI 48109, USA}
\email{liac@umich.edu}

\author[0000-0001-8470-749X]{Elisa Costantini}
\affiliation{SRON Netherlands Institute for Space Research, Leiden, The Netherlands}
\email{e.costantini@sron.nl}

% ...

\begin{abstract}

The \xrism\ \resolve\ microcalorimeter array measured the velocities of hot
intracluster gas at two positions in the Coma galaxy cluster: $3'\times3'$
squares at the center and at 6\am\ (170 kpc) to the south. We find 
the line-of-sight velocity dispersions in those regions to be 
$\sigma_z=208\pm12$ \kms\ and $202\pm24$ \kms,
respectively. The central value corresponds to a 3D Mach number of
$M=0.24\pm0.015$ and the ratio of the kinetic pressure of 
small-scale motions to thermal pressure
in the intracluster plasma of only $3.1\pm0.4$\%, at the lower end 
of predictions from cosmological simulations for merging clusters like 
Coma, and similar to that observed in the cool core of the relaxed cluster 
A2029. Meanwhile, the gas in both regions exhibits high line-of-sight 
velocity differences from the mean velocity of the cluster galaxies, 
$\Delta v_{z}=450\pm15$ \kms\ and $730\pm30$ \kms, respectively. 
A small contribution from an additional gas velocity component, consistent 
with the cluster optical mean, is detected along a sightline near the cluster
center. The combination of the observed velocity dispersions and bulk 
velocities is not described by a Kolmogorov velocity power spectrum of 
steady-state turbulence; instead, the data imply a much steeper effective
slope (i.e., relatively more power at larger linear scales). This may 
indicate either a very large dissipation scale resulting in the suppression
of small-scale motions, or a transient dynamic state of the cluster, where
large-scale gas flows generated by an ongoing merger have not yet cascaded
down to small scales.

\end{abstract}


\section{Introduction}

The weather in galaxy clusters is forecast to be stormy \citep{Burns98}.
X-ray images and temperature maps of the hot intracluster medium (ICM) have
long suggested that clusters are dynamic objects --- they show infalling
subclusters undergoing ram pressure stripping and accompanied by shock
fronts, as well as sharp ``cold fronts'' \citep[e.g.,][]{Jones99, Briel94,
Markevitch07}. The cluster cores often exhibit buoyant AGN
bubbles \citep[e.g.,][]{Churazov00, Fabian06, McNamara07}. Cluster radio
images frequently show radio galaxies with bent and contorted tails, 
believed to trace the ICM winds \citep[e.g.,][]{Burns98, Botteon20}.

The recently launched X-ray microcalorimeter array \resolve\ onboard the
\xrism\ observatory \citep{Tashiro20, Ishisaki22} picks up where its
short-lived twin, \hitomi\ SXS \citep{Takahashi16, Kelley16}, left off. 
These instruments provide the first precise, {\em direct}\/ look at
gas velocities in galaxy clusters across various dynamical states. A
microcalorimeter enables nondispersive high-resolution X-ray spectroscopy
(by recording the energy of each incident photon independently of its
source's position in the sky), allowing us to map plasma velocities in
clusters and other extended X-ray sources, such as supernova remnants, by
measuring the Doppler line shifts and broadening of the plasma emission
lines.

The line-of-sight (LOS) velocities and velocity dispersions have already
been reported for the cool cores of the Perseus \citep{Hitomi16}, Centaurus
\citep{xrismcen}, and A2029 \citep{xrisma2029} clusters. The dispersions are
found to be in the $\sigma_z\sim 120-190$ \kms\ range%
\footnote{Hereafter the index $z$\/ denotes the LOS component 
of the velocity.},
which constrains the mechanical energy output of the cluster's
central AGN and its contribution to the thermal balance of the cores. The
LOS velocities exhibit significant gradients across the cool cores in
Perseus and Centaurus, revealing gas sloshing likely triggered by 
past cluster mergers.

\begin{figure}
\centering
\vspace*{2mm}
\includegraphics[width=8.0cm]{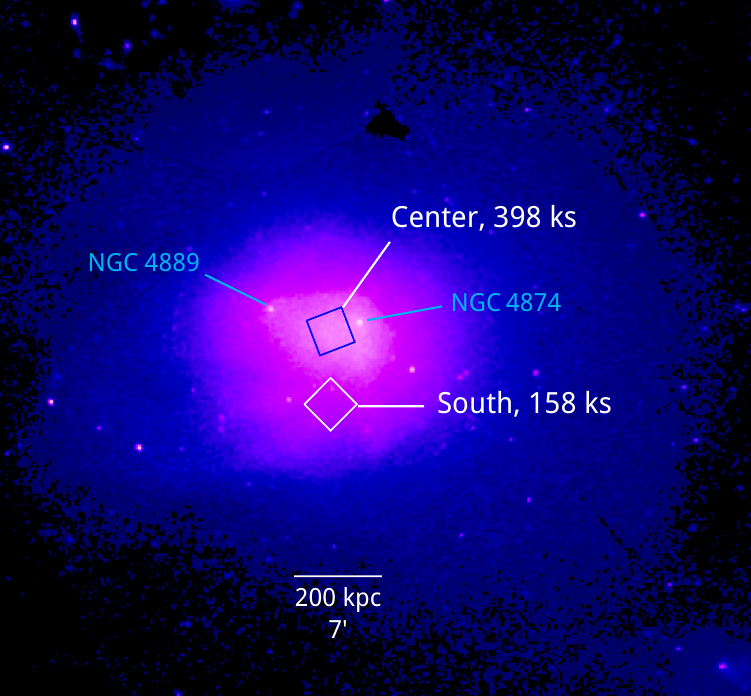}
\caption{\xrism\ Resolve fields of view and exposure times overlaid on the 
\xmm\ image of the Coma cluster \citep{Sanders20}. The two brightest cluster 
galaxies are marked.}
\label{fig:pointings}
\end{figure}

In this Letter, we present the first \resolve\ measurements for a disturbed
cluster without a cool core or a central powerful AGN --- the nearby
($z=0.02333$, \citealp{Bilton18}) Coma cluster. The cluster core has two
dominant galaxies (BCG), NGC\,4874 and NGC\,4889, with LOS velocities
differing by 720 \kms\ \citep[e.g.,][]{Colless96}. Both galaxies host 
small-scale X-ray coronae of $T=1-2$ keV gas with $r\sim$3 kpc --- likely
remnants of past cool cores stripped by ram pressure during a merger --- 
surviving within the hot $T=8-9$ keV ICM \citep{Vikhlinin01, Sanders14}. 
The galaxies also mark larger-scale bumps in the hot ICM density 
\citep{Vikhlinin97, Andradesantos13}, indicating underlying concentrations 
of dark mass, which are indeed seen in the weak-lensing mass map 
\citep{Okabe14}. 

On larger scales, the lensing map reveals distinct subhalos
\citep{Okabe14}. The ICM exhibits filamentary structures \citep{Vikhlinin97,
Sanders13}, possibly resulting from stripping of merging subclusters. 
A prominent shock front $\sim 1$ Mpc west of center is propagating in 
the sky plane, as well as a contact discontinuity at a similar distance 
to the east \citep{Planck13, Churazov21}, and a subgroup 1.3 Mpc to 
the southwest (outside the main X-ray cluster body) is apparently on a
return trajectory after passing through the main cluster
\citep{Churazov21}. All of the above indicates the presence of merging
activity in the plane of the sky.

\begin{table*}[tb]
\centering
\label{table:fits}
\renewcommand{\tabcolsep}{3mm}
\small\noindent
\caption{Parameters for one-component fits to spectra in full Center and South regions}
\hspace*{-1.5cm}
\begin{tabular}{p{3.2cm}ccccc}
\hline\hline
 & \multicolumn{2}{c}{Center}    & & \multicolumn{2}{c}{South}\\
 \cline{2-3} \cline{5-6}
 & 2--9 keV      & 6.4--6.9 keV   & &  2--9 keV  &  6.4--6.9 keV\\
\hline
$T$, keV\dotfill& $8.37\pm0.15$ & $8.55\pm0.25$& &$7.53\pm0.25$  &$7.44\pm0.44$\\
Fe abundance \dotfill& $0.32\pm0.01$&$0.33\pm0.02$& &$0.36\pm0.025$&$0.32\pm0.04$\\
$z$\dotfill   &         & $0.02183\pm0.00005$ & &          & $0.02089\pm0.00009$\\
$\sigma_z$, \kms\dotfill &    & $208\pm12$    & &          & $202\pm24$   \\
\hline
\end{tabular}
\end{table*}

Coma hosts a cluster-wide giant radio halo \citep{Brown11} --- the
synchrotron emission from ultrarelativistic electrons spinning in the
magnetic field permeating the ICM. These electrons are believed to be
continuously energized by turbulence in the ICM, though the efficiency of
this mechanism is uncertain \citep{Brunetti14}. Along with random
velocities, turbulence in the ICM should produce fluctuations in plasma
density and pressure, which have indeed been observed in Coma using
X-ray surface brightness and temperature maps \citep{Schuecker04, Churazov12,
Zhuravleva19, Sanders20}.

With relatively flat gas density and temperature profiles in the central
region \citep[e.g.,][]{Arnaud01, Sanders20} and thus no steep
radial entropy gradients (such as those present in 
cluster cool cores), as well as absence of AGN injecting bubbles into the ICM,
Coma offers perhaps the simplest experimental setup among galaxy clusters 
to study ICM turbulence. It should be driven solely by structure formation 
and develop in a simple, isotropic manner, free of the complications of a
stratified atmosphere. The goal of this work is to probe turbulence in the Coma
core using the first precise measurements of ICM velocities.

We use $H_0=70$ \kmsmpc, $\Omega_m=0.3$ flat cosmology, in which $1'=28.2$ 
kpc at the cluster redshift. The uncertainties are 68\%.

\section{Data}
\label{sec:data}

\subsection{Observations}

\xrism\ observed Coma during 2024 July 9-18 (obsid 300073010) with an 
aimpoint at $\alpha=194.944$\deg, $\delta=27.947$\deg\ near the cluster's 
X-ray center, referred to as ``Center'', and during 2024 May 20-24 
(obsids 300074010 and 300074020), with an aimpoint at 
$\alpha=194.941$\deg, $\delta=27.847$\deg, 6\am\ south of the center, 
referred to as ``South''. In this paper, we utilize data from the
\resolve\ instrument --- a microcalorimeter array that covers a 
$3.1'\times3.1'$ area of the sky with $6\times6$ pixels (except for
one corner pixel illuminated by an internal calibration source), 
each producing a spectrum of incident X-rays with a resolution 
of 4.5 eV FWHM \citep{Porter24}. The two \resolve\ pointings are overlaid 
on an \xmm\ image of the cluster in Fig.\ \ref{fig:pointings}.

The instrument's energy band spans $E=1.7-12$ keV (limited at low energies by
the attenuation of the window in the dewar gate valve that is currently
closed) and includes the Fe {\sc xxv-xxvi} emission line complex at
$E=6.7-6.9$ keV (rest-frame), a dominant feature in the spectrum of the hot,
optically-thin ICM.

\subsection{Data Reduction}

We extract spectra from the \resolve\ photon lists produced by the 
\xrism\ pipeline (Build 8, CalDB version 8 (20240315)), 
following the procedure detailed in \citet{xrisma2029}. We use 
only the high-resolution primary events. The \resolve\ pixel 27, which 
exhibits poorly modeled gain excursions, is excluded from the spectra, 
along with calibration pixel 12. The standard screening yields clean
exposures of 398 ks for the Center and 158 ks for the South (where we 
co-add the two coaligned partial exposures of 85 ks and 73 ks). The 
heliocentric velocity corrections, accounting for the Earth's velocity 
component toward the target, are --23.5\kms\ (Center) 
and --21.6\kms\ (South); all velocities and redshifts below are given 
in the heliocentric frame. We use the spectral redistribution matrix (RMF)
of ``L'' size%
\footnote{heasarc.gsfc.nasa.gov/docs/xrism/analysis/abc\_guide/xrism\_abc.html
\label{fn:abc}}
for the
results below; no significant changes to the results were found between
``M'', ``L'' or ``X'' size matrices (which differ in the tradeoff between
accuracy of modeling of secondary response components and convolution speed).

The \resolve\ energy scale (gain) is continuously calibrated in orbit,
resulting in an energy scale uncertainty of $\leq0.3$ eV (field averaged) for 
the 5.4--9 keV band \citep{Eckart24, Porter24, Eckart25}, corresponding to
$\leq 15$\kms\ Doppler shift instrumental uncertainty for a line at 6 keV. 

The \resolve\ charged-particle induced non-X-ray background (NXB) is
negligible for deriving the shapes of the cluster Fe lines, but it reaches
about 10\% of the continuum signal from the Coma core at both ends of the
useful band (at $E\sim 2$ keV and 9 keV), so it needs to be accounted for
when modeling the continuum. The cosmic X-ray background (CXB) is $\lax 1$\%
of the cluster signal anywhere in this band. Therefore, for the spectral
fits below, we included the NXB spectral model (continuum plus several
emission lines) fit to the \resolve\ blank-sky data as described in
\citet{xrismcen}, but disregarded CXB for simplicity.

The angular resolution of the X-ray mirror is 1.3\am\ (half-power diameter, 
HPD, of the point-spread function, PSF). For spectra extracted from the
full 3\am\ \resolve\ field of view (FOV), intermixing of photons from 
adjacent regions in the sky is relatively minor, especially for a cluster
like Coma with a non-peaked X-ray brightness distribution. However, in
spectra extracted from 1.5\am\ quadrants, about 50\% of the recorded 
photons are scattered from adjacent regions. For qualitative purposes of
\S\ref{sec:t}--\S\ref{sec:powspec}, we use spectral fits derived by
disregarding PSF scattering (fitting spectra from adjacent regions
independently of each other and using ancillary response functions, ARF,
generated for point sources%
$^{2}$%
), so those values are approximate. 
We do include PSF mixing in the forward modeling of the velocity differences
in \S\ref{sec:powspec}, where it has a significant effect, ensuring our results
for the velocity power spectrum are precise.

\begin{figure*}
\centering
\includegraphics[width=17.5cm]{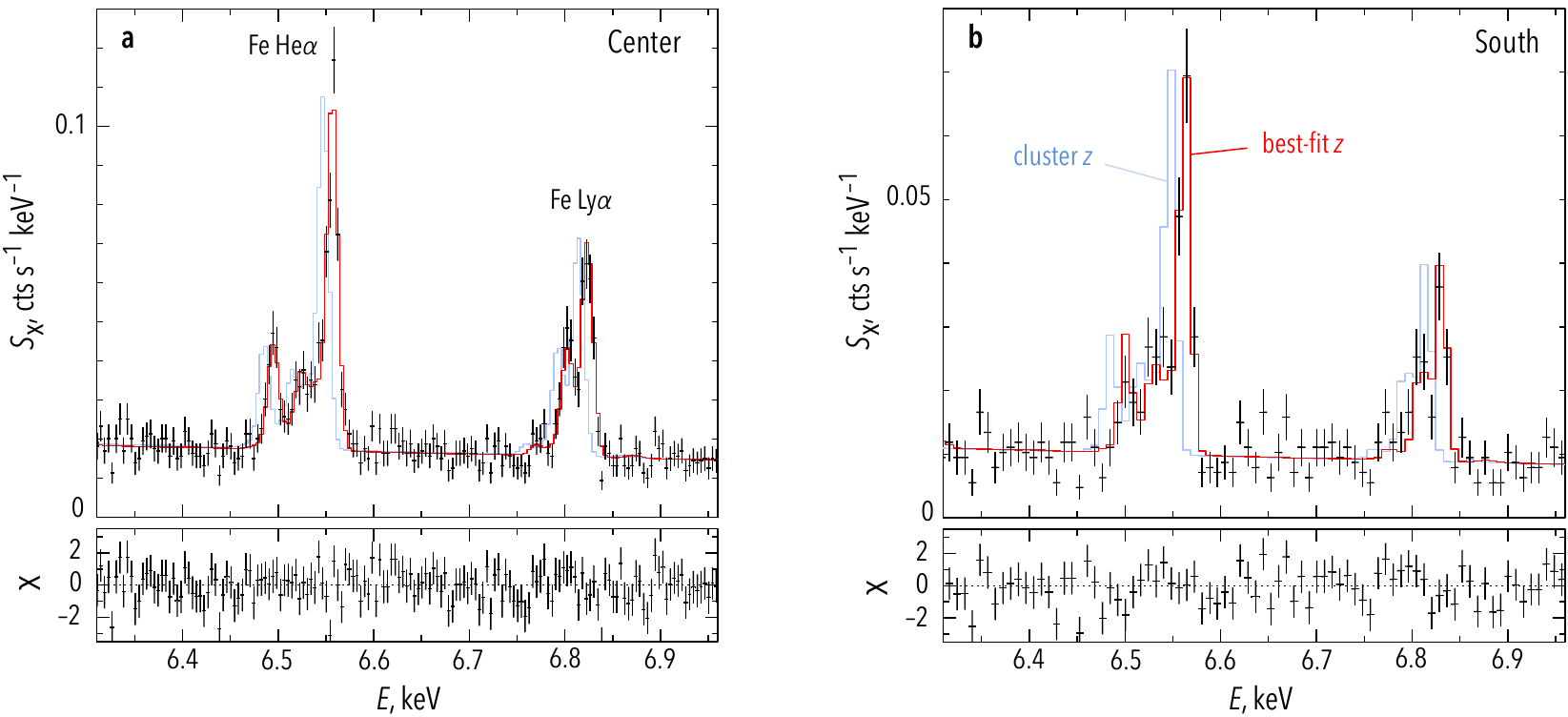}

\caption{\xrism\ \resolve\ spectra for (a) Center and (b) South pointings in
the He-like and H-like Fe line region, binned by 4 eV and 8 eV,
respectively. Red lines show the best-fit models (Table
\ref{table:fits}), while the light blue line represents a model with the
cluster optical redshift; a shift is evident.}
\label{fig:spec}
\end{figure*}

\section{Results}

\subsection{Gas temperatures and abundances}
\label{sec:t}

Because Coma lacks a cluster-scale cool core and its 
projected temperature across the core is relatively uniform 
\citep[e.g.,][]{Arnaud01, Sanders20},
we expect the gas within each of our pointings to be well described by a
single-temperature model. We therefore fit the \resolve\ spectra from the
entire Center and South FOV (using the {\sc xspec} package, \citealp{xspec}) 
with a one-component thermal plasma emission
model that includes thermal broadening ({\sc bapec}, \citealp{Smith01}),
abundances relative to solar from \cite{Asplund09}, fixing Galactic 
absorption at $N_H=9.2\times 10^{19}$~\cmsq\ (which is unimportant for 
our energy band), and using a broad 2--9 keV band along with a narrow 
6.4--6.9 keV interval that encompasses the Fe {\sc xxv-xxvi} complex. 
The spectra for both regions are well-fitted, free of any systematic
residuals; the fit parameters are provided in Table\ \ref{table:fits}.
Importantly, we find that the broad-band temperatures, primarily 
determined by the continuum slope, and the narrow-band 
temperatures, primarily determined by the Fe {\sc xxvi/xxv}
flux ratio, are in good agreement within their tight errors. This gives
confidence in the accuracy of the temperatures and the gas velocity
dispersions (below), whose effect on the Fe line width combines (in quadrature)
with the thermal broadening ($\sigma_{{\rm th},z}=120$ \kms\ and 112 \kms\ for 
the Center and South temperatures, respectively). These temperatures will be
compared with those from other X-ray instruments in a forthcoming paper. 
Based on these temperatures, the estimated sound speeds are 
$c_s=(\gamma k T/\mu m_p)^{1/2}=1508$\,\kms\ and 1407\,\kms\ for the Center 
and South regions, respectively (where $\gamma=5/3$ is the polytropic index
and $\mu=0.6$ is the mean molecular weight of the intracluster plasma).

We also fit the Center spectrum with a model with independent abundances 
({\sc bvapec}). In addition to Fe, other elements with notable abundance
constraints include Ni ($0.43\pm0.11$), S ($0.27\pm0.12$) and Ar
($0.49\pm0.21$); others are detected at $<2\sigma$ significance.

\begin{figure*}
\centering
\includegraphics[width=8.3cm]{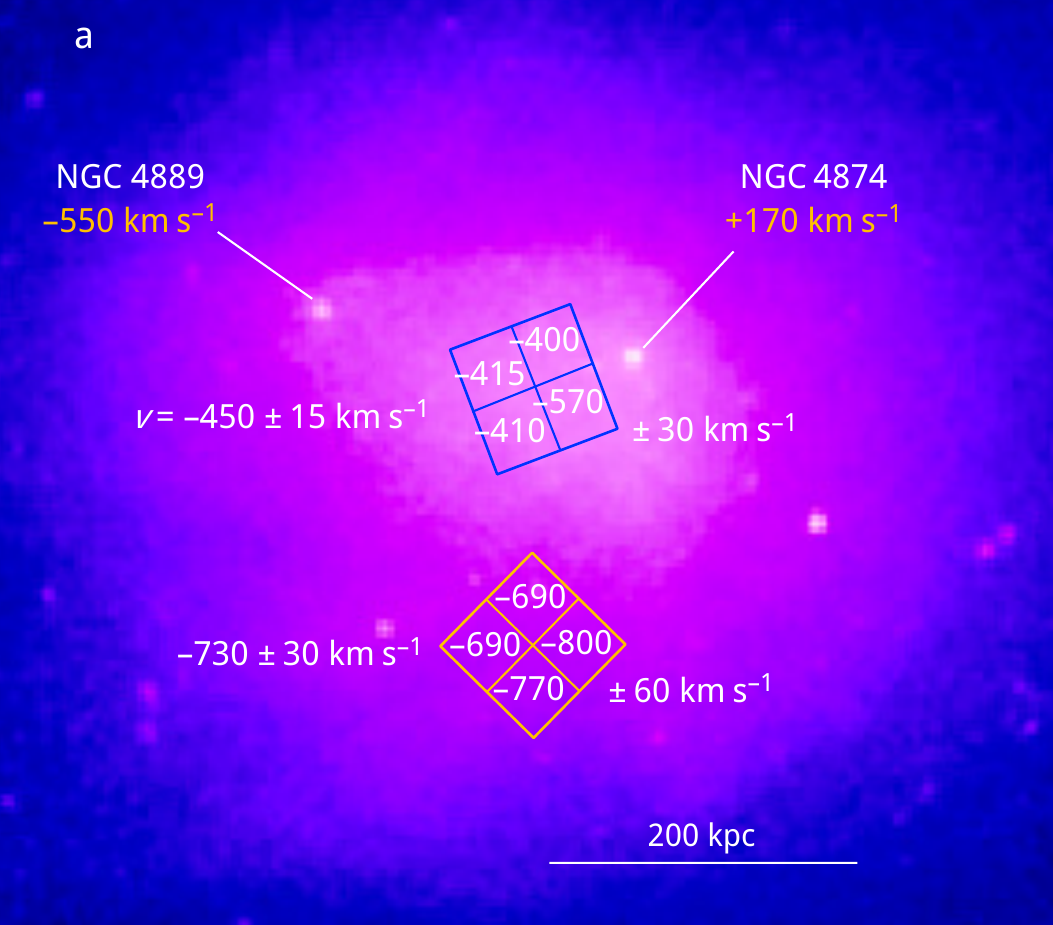}
\hspace*{1mm}
\includegraphics[width=8.3cm]{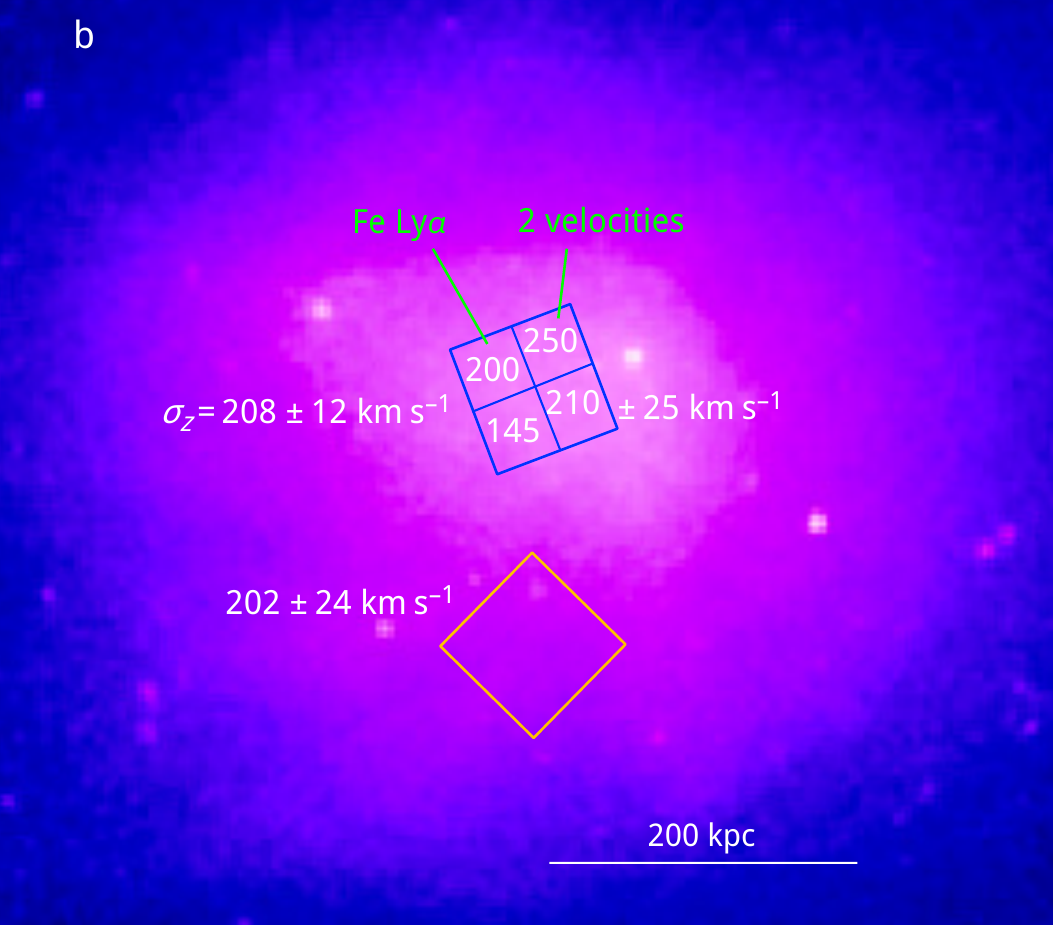}

\caption{\xrism\ \resolve\ measurements of ({\em a}) the LOS velocities relative
to the mean velocity of cluster member galaxies and ({\em b}) LOS velocity
dispersion, overlaid on the \xmm\ image. Uncertainties are statistical 
$1\sigma$. The values to the left of the fields represent the entire 
$3'\times3'$ field, while the values inside the FOV pertain to the 1.5\am\ 
quadrants, for which the relatively small PSF smearing effect is 
not included. The two brightest galaxies are marked along with their 
relative LOS velocities. Green labels mark the quadrants
discussed in \S\S\ref{sec:vels}--\ref{sec:lya}.
} 
\label{fig:vels}
\end{figure*}

\begin{figure}
\centering
\vspace*{2mm}
\includegraphics[height=7.6cm]{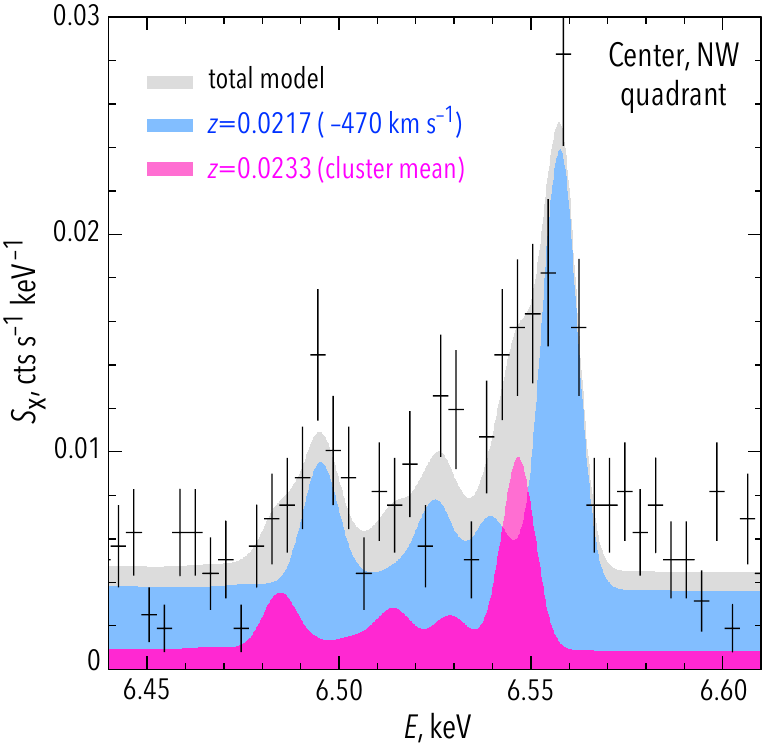}

\caption{The spectrum for the NW quadrant of the Center field (the one with
the highest velocity dispersion, see Fig.\ \ref{fig:vels}{\em b}). Only the 
Fe-He$\alpha$ line complex is displayed, binned by 4 eV. Two velocity
components are needed to model the line profile. The main component (blue) 
is blueshifted from the cluster mean by 470 \kms, while the additional
component (magenta) is at the cluster optical mean velocity and contributes
$22\pm7$\% of the emission measure.}  
\label{fig:2comp}
\end{figure}

\subsection{Gas velocities and dispersions}
\label{sec:vels}

The spectra of the Fe complex are shown in Fig.\ \ref{fig:spec}. The
best-fit redshifts and LOS velocity dispersions $\sigma_z$\/ with their 
statistical uncertainties are presented in Table \ref{table:fits}. The 
lines in both regions (a) are narrow, with $\sigma_z\approx200$\kms, and 
(b) show large velocity offsets
from the cluster optical redshift ($cz=6995\pm39$ \kms, \citealp{Bilton18}),
$\Delta cz=-450\pm15$\kms\ for the Center and $-730\pm30$\kms\ for the South.
These offsets are evident in Fig.\ \ref{fig:spec}; they correspond to line 
shifts of 10 eV and 16 eV, much greater than the gain calibration uncertainty
of 0.3 eV (\S\ref{sec:data}). Both velocity offsets align with
those derived in larger regions with \xmm\ \citep{Sanders20}, within the
latter's 10--20 times larger uncertainties.

The Center spectrum has approximately 940 and 510 counts in the 
Fe{\sc xxv} and {\sc xxvi} complexes, respectively, which is sufficient for
deriving velocities and dispersions in separate quadrants with good
statistical precision. In contrast, the South spectrum has a total of 480 
line counts. Results for the quadrants in the Center pointing, along
with their typical uncertainties, are shown in Fig.\ \ref{fig:vels} 
(except for dispersions in the South, which have large uncertainties). 
There are significant variations in line width within the Center; the SE
quadrant shows a narrower line than the field average, 
while the NW quadrant exhibits a broader line. 

The Fe {\sc xxv} He-$\alpha$ complex for the NW quadrant is shown in
Fig.\ \ref{fig:2comp}; the line shape suggests the presence of at least one
additional component at an energy below the main peak. Allowing for two
plasma components with free $z$\/ and $kT$\/ and the same (free) velocity
dispersion and chemical abundances, improves the fit by $\Delta \chi^2=10.6$
for 3 additional parameters.  This component contributes $22\pm7$\% of the
total emission measure, with a best-fit $z=0.0233\pm0.0004$ consistent with
the cluster's optical redshift, while the main component deviates from it by
$\Delta cz=-470$\kms. Their best-fit dispersion is $127\pm41$ \kms, with
temperatures of $9.4\pm0.8$ keV and $6.8\pm1.8$ keV for the main and
additional components, respectively.  While temperatures of the two
components are statistically indistinguishable, the relatively high
temperature for the additional component does suggest that the gas is
located in the cluster core (rather than being projected from the cooler
outskirts) --- possibly related to the X-ray brightness excess associated
with NGC\,4874 \citep{Vikhlinin97, Andradesantos13}.

\subsection{Fe Ly$\alpha$ anomaly}
\label{sec:lya}

\begin{figure}[t]
\centering
\vspace*{2.0mm}
\includegraphics[height=7.55cm]{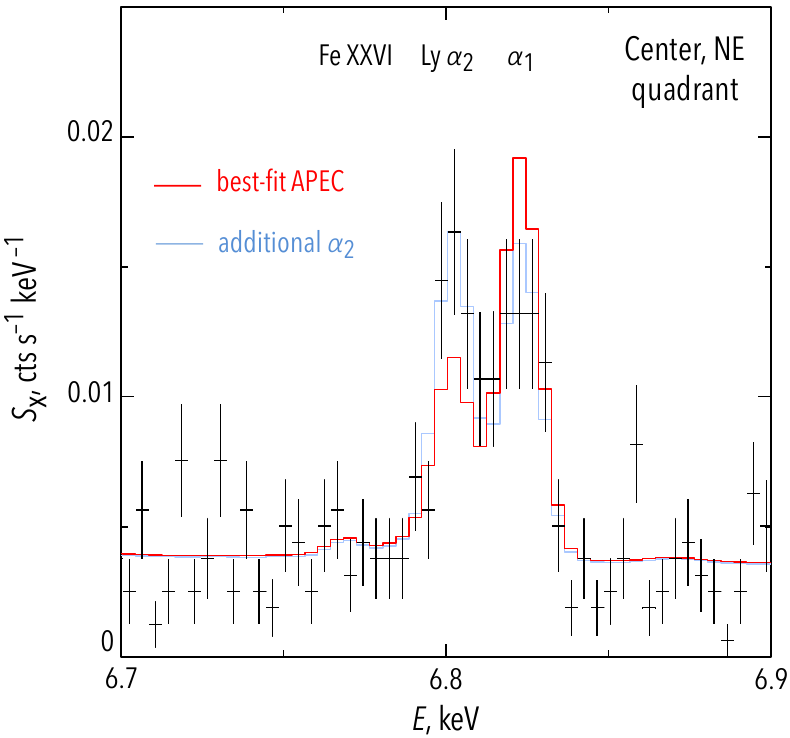}

\caption{The spectrum from the NE quadrant of the Center field around the
Fe-Ly$\alpha$ line, binned by 4 eV. The red line shows a {\sc bapec} model fit
in the interval shown in Fig.\ \ref{fig:spec} (including Fe He$\alpha$ and
Ly$\alpha$ lines). An excess is observed in the Ly$\alpha_2$ component. The 
blue line shows a fit with additional flux in this line, required at
$\sim3\sigma$ significance.}
\label{fig:lya}
\end{figure}

A detailed look at the spectra in the Center quadrants reveals an apparent
excess of the Fe Ly$\alpha_2$ line flux over the model in the NE quadrant
(Fig.\ \ref{fig:lya}). The other three quadrants, or the South 
region (Fig.\ \ref{fig:spec}{\em b}), do not show such anomaly.
An additional flux in this line component is required at a 
$\sim3\sigma$ significance;
given that we searched 5 spectra for this signal (four Center 
quadrants and South), this corresponds to a detection at a $\sim98$\%
statistical confidence level. The line cannot be attributed to an additional
velocity component. Resonant scattering in the Fe Ly$\alpha$ lines, 
proposed to explain a modified $\alpha_1/\alpha_2$ 
flux ratio (compared to the theoretically expected 2:1) 
in some other sources \citep[e.g.,][]{Gunasekera24}, has negligible
optical depth in Coma \citep[e.g.,][]{Sazonov02}.
There is nothing anomalous at this location in the X-ray
(\chandra\ or \xmm), optical, or radio images of the cluster. Interestingly,
the spectrum of A2029 \citep{xrisma2029} exhibits a similar anomaly,
although with lower amplitude. Physical explanations for this anomaly 
will be explored in future work.

\section{Discussion}

\subsection{Velocity dispersion and kinetic pressure}

Assuming isotropic velocities, the observed LOS velocity dispersions
(Table \ref{table:fits}) correspond to Mach numbers of 
$M_{\rm 3D}=\sqrt{3} M_z = 0.24\pm0.015$ and $0.25\pm0.03$ for the 
small-scale gas motions in the Center and South pointings, respectively. 
The fraction of kinetic pressure in total ICM pressure, estimated as
\citep[e.g.,][]{Eckert19}
\begin{equation}
\frac{p_{\rm kin}}{p_{\rm tot}}=
\left(1+\frac{3}{\gamma M_{\rm 3D}^2}\right)^{-1},
\label{eq:pkin}
\end{equation}
is $3.1\pm0.4$\% and $3.3\pm0.8$\% for the Center and South
sightlines. This estimate does not include pressure contributions from 
bulk velocities, as it is based on local dispersions relative to the local
bulk velocities in the C and S fields; however, it does use the full 
velocity dispersion without attempting to separate
large-scale and small-scale variations along the LOS. Notably, 
this pressure ratio is similar to 
$p_{\rm kin}/p_{\rm tot}=2.6\pm0.3$\% that \xrism\ measured (in a 
similar manner) in the cool core of A2029 \citep{xrisma2029}, one of
the most relaxed clusters known.

\begin{figure}[tb]
\centering
\vspace*{2mm}
\includegraphics[width=7.5cm]{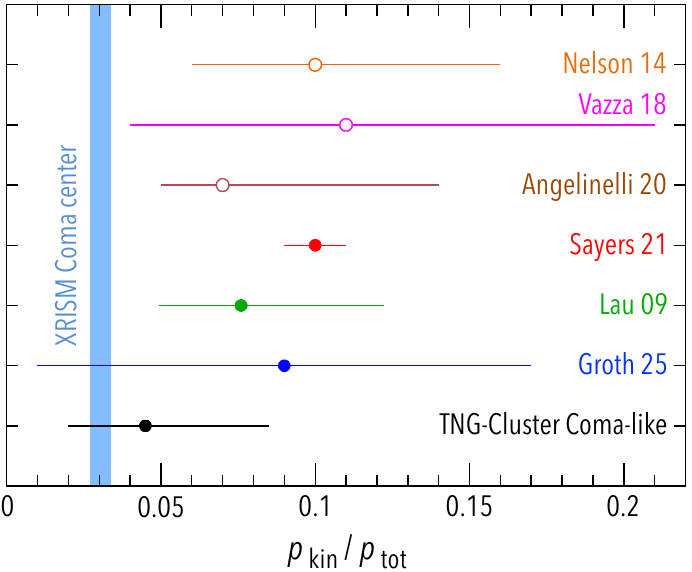}

\caption{The \xrism\ value of $p_{\rm kin}/p_{\rm tot}$ for Coma, derived from
the central line width, is compared with predictions from several cosmological
simulations. Open symbols represent all clusters, while filled symbols
select only the disturbed Coma-like clusters. For TNG-Cluster, there is an
additional selection based on cluster mass and weighting by X-ray emission 
measure (see Appendix B). The horizontal intervals show the 
$16^{\rm th}-84^{\rm th}$ percentiles of the simulated cluster
populations. The measured value is at the low end of the predictions.}
\label{fig:pturb}
\end{figure}

Figure \ref{fig:pturb} compares this kinetic pressure in the
Coma Center to predictions from several cosmological simulations, selecting
sightlines through the cluster center and estimating $p_{\rm kin}$ based on
the LOS velocity dispersions, as done for Coma, while choosing Coma-like
disturbed clusters whenever possible (see Appendix \ref{app:cosmosim}).

Our measured velocity dispersion is at the lower end of, or in many cases
lower than, the simulation predictions. One might expect that the finite
numerical resolution in cosmological simulations would lead to an
underestimation of ICM velocities on small linear scales; however, we
observe the opposite. The dispersion in the cool core of A2029
is also found to be on the low end of predictions \citep{xrisma2029}. 
Upcoming \xrism\ observations of other clusters will determine whether
this is more than a statistical fluctuation.

\subsection{Velocities of gas versus galaxies}
\label{sec:gals}

\begin{figure}[tb]
\centering
\vspace*{2mm}
\includegraphics[width=8.2cm]{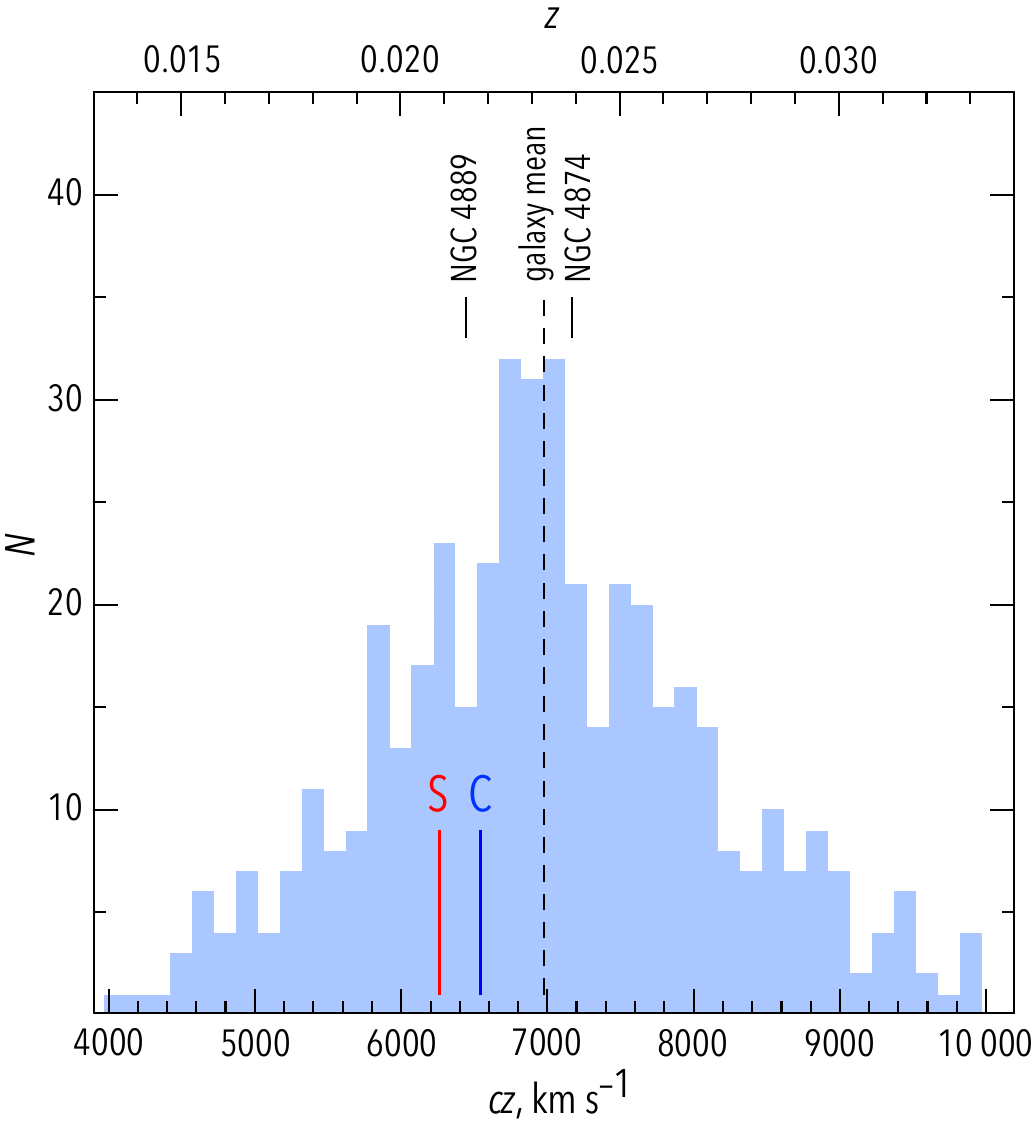}
\caption{A histogram of velocities of cluster member galaxies within $r<20'$
from the Coma X-ray center, retrieved from SDSS, DESI, and \cite{NED}
archival data.  The gas velocities measured with \xrism\ for the Center and
South fields (marked C and S, respectively) are offset from the cluster
galaxy mean.  The two main BCGs are marked; NGC\,4874 is located near the
Center field.}
\label{fig:hist}
\end{figure}

The galaxy velocity distribution in the central $r<20'\sim 0.6$ Mpc 
region of the cluster (which excludes the well-separated infalling group
centered on NGC\,4839) is well described by a single Gaussian 
(\citealp{Colless96}; see our updated velocity histogram in Fig.\ 
\ref{fig:hist}). The two BCGs exhibit a large LOS velocity difference, 
and several small subgroups are detected in the ($\alpha, \delta, z$)
distribution of the cluster galaxies in the cluster central region 
\citep[e.g.,][]{Healy21}, but no 
prominent substructure indicates an ongoing major merger along the line
of sight. It is therefore puzzling why the gas has such a large LOS velocity
offset from the cluster galaxies.  The velocities in both \xrism\ pointings
are closer to that of the secondary BCG, NGC\,4889 (Fig.\ \ref{fig:vels}{\em a}),
rather than the main BCG NGC\,4874 (which is nearer to the X-ray
centroid, the galaxy velocity mean, and the main peak of the mass map in
\citealp{Okabe14}), as previously suggested by \cite{Sanders20}. As noted
above (Fig.\ \ref{fig:2comp}), there is only a small
amount of gas along the sight line near NGC\,4874 that matches that galaxy's 
velocity. Contrary to the expected scenario of gas and galaxies in
approximate hydrostatic equilibrium, our measurements indicate a wind of gas
with low random internal motions ($M_{\rm 1D}=0.14$) flowing through the
galaxies at a relatively high speed, $M_{\rm 1D}=0.3-0.5$. This
picture is based on data from only two locations in the core and may 
change as more sight lines are sampled by \xrism.

\begin{figure*}[tb]
\centering
\includegraphics[height=8.45cm]{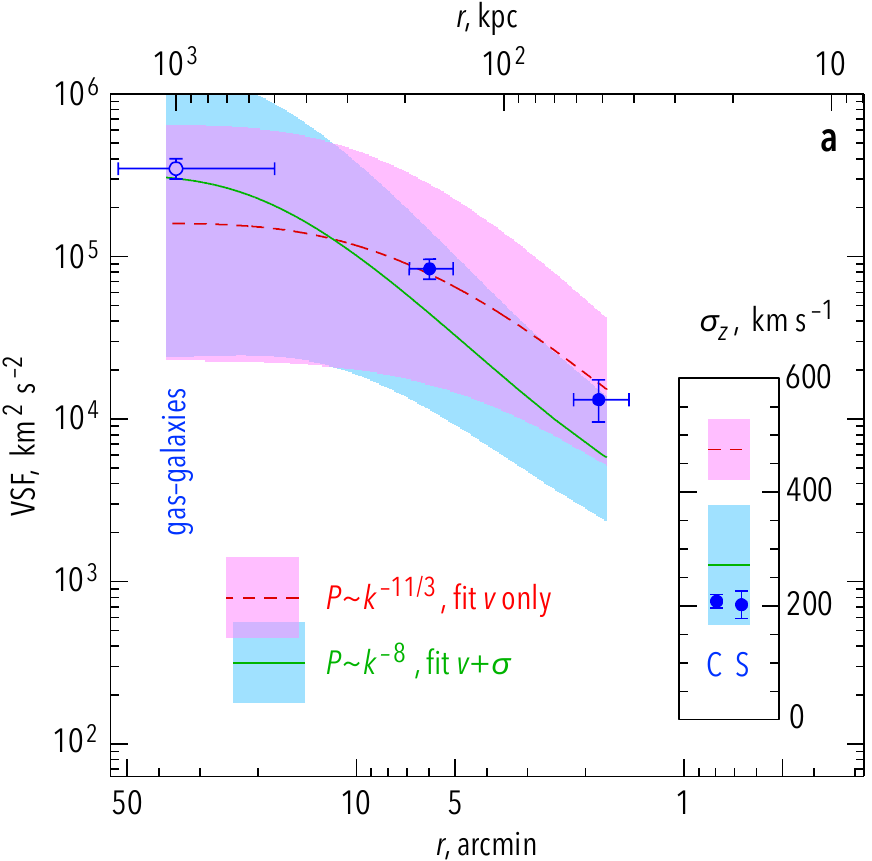}
\hspace*{8mm}
\includegraphics[height=8.4cm,trim=0 1.0mm 0 0]{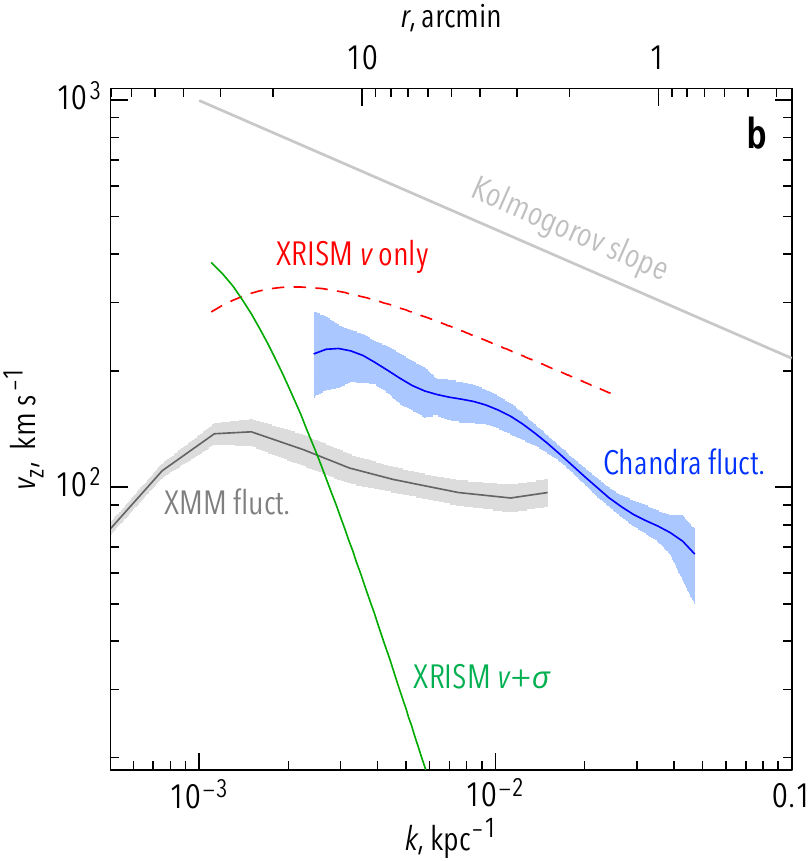}

\caption{Turbulence power spectrum inferred from the observed gas velocity
differences and dispersions. ({\em a}) The velocity structure function is 
shown by blue crosses. The two velocity dispersions are shown in the inset 
by blue symbols. The gas-galaxy offset (see text) is included as a 
velocity difference at 1 Mpc (open symbol). The red dashed line fits  
a Kolmogorov spectrum with $\ell_{\rm inj}$=1 Mpc and $\ell_{\rm dis}$=1 kpc
to the VSF points only; it significantly overestimates the line broadening.
The green line shows a model with a steeper slope that fits all \xrism\ data 
(velocity differences and dispersions). The color error bands for the models
include the 68\% cosmic variance uncertainty and the
measurement statistical errors (the latter are also shown as error bars on
data points for illustration). The horizontal bars show approximate bin
sizes. ({\em b}) Velocity amplitudes for the same two power spectra. Spectra
from X-ray surface brightness fluctuation analyses of \chandra\ and
\xmm\ images are shown for comparison \citep{Zhuravleva19, Sanders20} (note
these are not directly comparable; see text).}
\label{fig:powerspec}
\end{figure*}

\subsection{Velocity power spectrum}
\label{sec:powspec}

The ICM is primarily heated by thermalizing the kinetic energy of gas released
during cluster mergers and matter infall, which create shocks and
turbulence. The physics of this energy conversion is encoded in the amplitude 
of the cluster gas velocity variations, ${\bf v}({\bf x})$, as a function 
of the length scale $\ell$ (or wavenumber $k\equiv 1/\ell$). This can be 
quantified by a power spectrum $P(k)\equiv |\tilde{v}_z({\bf k})|^2$, 
where $\tilde{v}_z$ is the Fourier transform of one component of the 
velocity field $v_z$, ${\bf k}$ is the 3D wavevector, and (assuming 
isotropy of the power spectrum of the velocity field) $k\equiv |{\bf k}|$.

Gas motions are injected on a 
large linear scale $\ell_{\rm inj}$ by mechanical disturbances (such as 
cluster collisions, buoyant AGN bubbles, or random galaxy motions); 
these motions randomize, generate velocity variations on smaller scales, 
and dissipate (mostly into heat) at a scale $\ell_{\rm dis}$ determined 
by the microphysics of the ICM. In a steady state, the power spectrum 
attains a characteristic shape that can be described by:
\begin{equation}
P(k)  = P_0\;\left[1+(k\ell_{\rm inj})^2\right]^{\alpha/2}\;%
\exp\left[-(k\ell_{\rm dis})^2\right],
\label{eq:pk}
\end{equation}
where $P_0$ is the normalization. In the \cite{Kolmogorov1941} model, the
spectrum in the inertial range (between $\ell_{\rm inj}$ and $\ell_{\rm
dis}$) has a power law slope $\alpha=-11/3$, for which the flux of kinetic
energy down the cascade is independent of the scale.

With two \xrism\ pointings 6\am\ apart, whose spectral line positions
provide average LOS velocities in
1.5\am\ quadrants and whose line widths average the LOS velocity variations
over a range of scales (including the smallest ones that \xrism\ cannot
resolve in the sky plane), we can constrain $P(k)$. This
can be achieved by constructing a velocity structure function (VSF), which
quantifies the LOS velocity variation as a function of separation in the sky
plane and is directly related to $P(k)$ (Appendix A). The differences
between velocities in the Center quadrants (Fig.\ \ref{fig:vels}) are used
to evaluate ${\rm VSF}(r)$ at $r\sim 1.8'$. The velocity for the entire South
pointing is subtracted from each Center quadrant to construct another VSF 
bin at $r\sim 6'$ (statistical uncertainties in the South quadrants
are twice as large as those in the Center, so we chose not to use them.) 

For this exploratory exercise, we treat the large gas velocities
relative to the galaxies in our two pointings as another velocity difference
at a larger scale. We can reasonably assume that the gas and galaxies
fill the same potential well and have the same cluster-averaged velocities,
and use the galaxy average as a substitute for the gas average. We therefore
use the mean of the S and C velocity offsets from the average of cluster
galaxies, $-590$ \kms, to construct a VSF point at $r=1\pm0.5$ Mpc ($r=35'$); 
given all the uncertainties, the exact linear scale is not critical. The 
resulting VSF is shown in Fig.\ \ref{fig:powerspec}{\em a}, along with 
the dispersion values for the two pointings.

We model these data with a velocity power spectrum (eq. \ref{eq:pk}) as
described in Appendix A. Key details are: (a) We do not separate
turbulence from bulk motions \citep[as in, e.g,][]{Vazza12}; our
eq.\ (\ref{eq:pk}) describes the total velocity field. (b) Model
velocities are weighted by the cluster 3D X-ray emission measure
distribution, for which we use a symmetric $\beta$-model \citep{Briel92} for
simplicity. (c) Because the two \xrism\ pointings sample the cluster sparsely,
each VSF linear scale is probed by at most a few velocity differences. It is 
possible that they are outliers, but we have to treat them as representative
of the mean of the random velocity field. Additionally, the largest scales
of interest are comparable to the size of the cluster, so regardless of 
\xrism\ coverage, there may only be a couple of large eddies in the entire
cluster. These considerations lead to a large, but quantifiable,
``cosmic variance'' uncertainty in our modeling (\citealp{ZuHone2016};
Appendix A). Currently, this is the dominant uncertainty limiting our
conclusions.

We fix $\ell_{\rm inj}=1$ Mpc (a logical first guess for a merger) and
$\ell_{\rm dis}=1$ kpc (essentially zero, as it is well below
\xrism\ resolution) and fit the VSF data with a Kolmogorov ($\alpha=-11/3$)
model, with normalization as the only free parameter. This model is 
shown by the red dashed line in Fig.\ \ref{fig:powerspec}{\em a}, while the
red band combines the 68\% statistical uncertainty of the data and the
cosmic variance for the model into a single uncertainty interval. The 
model fits the {\em velocities}\/ well ($\chi^2=0.2$ for 2 d.o.f.); 
however, it predicts a dispersion, $\sigma_z=475\pm53$ \kms, 
far exceeding the observed line widths (Fig.\ \ref{fig:powerspec}{\em a}\/ 
inset). If we fit the VSF and dispersions together using the Kolmogorov 
slope, we obtain a poor fit with $\chi^2=13.0$ for 4 d.o.f. --- there 
is a clear tension between the velocities and the
dispersions, indicating a need for a steeper $P(k)$ model. The green
line with the blue band in Fig.\ \ref{fig:powerspec} shows one such model, 
with $\alpha=-8$ and the same $\ell_{\rm inj}$ and $\ell_{\rm dis}$. 
It fits better with $\chi^2=1.6$ for 3 d.o.f.; the 
F-test indicates a 98\% confidence that the slope must be steeper than 
Kolmogorov. The 68\% constraint on the slope is $\alpha<-4.8$. 

Another way to steepen the spectrum is to use a large value for 
$\ell_{\rm dis}$; an indistinguishable fit to the VSF and dispersions is 
achieved for $\ell_{\rm dis}=1000$ kpc, or $\ell_{\rm dis}>240$ kpc at 68\% 
confidence. (The best-fit values for $\alpha$ and $\ell_{\rm dis}$ 
are merely the upper bounds of our trials, where the goodness of fit
depends only weakly on the exact value of the parameter, and so they do 
not carry as much physical significance as their 68\% bounds.)

We note that the LOS velocity dispersion incorporates contributions from 
all linear scales (weighted by the X-ray emission along the LOS), not just 
from small scales. When fit to the velocity and dispersion data, our 
steeper-spectrum models are more consistent with the low observed 
dispersions largely because the predicted scatter of $\sigma_z$ among 
different sightlines (the cosmic variance) is higher for steep-spectrum models
(Fig.\ \ref{fig:powerspec}{\em a}\/ inset), making it less improbable to
encounter random deviations as low as the observed $\sigma_z$. If future
pointings yield similarly low dispersions, this model will become less tenable.

Velocity amplitudes for the Kolmogorov fit to the VSF and the steep-slope 
fit to all data are shown in Fig.\ \ref{fig:powerspec}{\em b}\/ (the 
high-$\ell_{\rm dis}$ model is qualitatively similar to the steep-slope 
one and is not shown). The models exhibit similar amplitudes around 
$\ell\sim 500$ kpc sampled by the bulk velocities but diverge at smaller
scales. The figure also shows two indirect estimates of the velocity 
variations derived from X-ray surface brightness fluctuations in 
\chandra\ and \xmm\ images \citep{Churazov12, Zhuravleva19, Sanders20},
assuming that the velocity variations are proportional to the
density fluctuations on scale $k$: $\delta\rho_k/\rho=\eta\, v_{z,k}/c_s$
with $\eta\approx1$ \citep{Zhuravleva14, Zhuravleva23}. The \chandra\ result
pertains to the Coma core and can be approximately compared with 
\xrism\ measurements in the core (although \xrism\ does not sample the 
entire \chandra\ region). The \chandra\ spectrum, when integrated along the 
LOS weighted with the X-ray emissivity (similarly to our $P(k)$ modeling, 
see Appendix A), would yield $\sigma_z=300$ \kms, a factor $\sim1.5$ 
higher than the \xrism\ dispersion measurements, which falls within the large 
uncertainties of this indirect method expected for merging clusters 
\citep{Zhuravleva23}. The \xmm\ fluctuation spectrum does not exceed 
the observed $\sigma_z$, although the \xmm\ spectrum is derived for the 
entire cluster ($r=1.5$ Mpc) and is not directly comparable to the \xrism\
measurements in the core.

The \chandra\ finding that the Kolmogorov slope extends to high $k$\/ has 
been interpreted as evidence for a very low effective isotropic viscosity
in the ICM \citep{Zhuravleva19}. The shape of our preferred $P(k)$ model
is qualitatively very different from both fluctuation-based results. 
However, with the available sparse data, we could only consider the 
simplest family of spectra (eq.\ \ref{eq:pk}) and cannot rule out the
possibility that a more complex shape would fit all \xrism\ data and 
be more similar in shape --- even if not in amplitude --- 
to the fluctuation results at higher $k$. It is thus crucial 
to validate the fluctuations-based spectral shape with
\xrism\ measurements in the range of scales where the instruments overlap
($1.5'-15'$ or $k=0.002-0.02$) and to confirm that the X-ray surface
brightness fluctuations indeed return the ICM velocities. This would require
a more complete coverage of the cluster core with multiple \xrism\ pointings,
aimed at (a) sampling more scales and (b) reducing the cosmic
variance uncertainties for the VSF and line widths.

\subsection{Velocity spectrum interpretation}
\label{sec:specint}

While the exact shape of the velocity power spectrum is still uncertain, it
is clear from \xrism\ results that it is steep, exhibiting relatively lower
velocities at small scales compared to a steady-state Kolmogorov model of
turbulence. Two broad possibilities exhist: (a) fast dissipation of motions
on intermediate linear scales, preventing power from reaching small scales,
or (b) a transient state, where random motions generated at large scales
have not yet cascaded down to small scales. The first possibility would
require a dissipation scale, $\ell_{\rm dis}\gax 240$ kpc, much larger than
the Coulomb collision mean free path in the Coma core, $\lambda\approx 10$
kpc, and consequently a very high effective viscosity of the ICM. The second
possibility requires an ongoing major merger, as the
velocities should cascade on a timescale of order the large eddy
turnover time, $2-3$ Gyr. However, the galaxy velocity distribution does
not indicate a strong merger (see \S\ref{sec:gals}). Both possibilities
have challenges but remain open for now, pending better constraints on the
shape of the spectrum.

The notion that cosmic ray electrons responsible for giant radio halos 
in clusters (including Coma) are reaccelerated by ICM turbulence depends on 
the turbulence spectrum extending to very small scales, where acceleration
occurs \citep{Brunetti14}. If motions dissipate at large scales, as happens 
in one of our possibilities, this explanation may become problematic. This 
underscores the importance of accurately determining the shape of the Coma
velocity spectrum.

\section{Summary}

Using \xrism\ \resolve, we conducted exploratory measurements of ICM
velocities in two $3'\times3'$ regions of the Coma cluster. We
detected a high-velocity wind blowing through the galaxies, but the velocity
dispersion is relatively low in both regions, implying a velocity power
spectrum with an effective slope steeper than that in the classic
Kolmogorov model of steady-state turbulence. This raises questions about the
dynamical state of the cluster and the physics of the ICM. Answers may 
be found in a more precise shape of the power spectrum.

\begin{acknowledgments}
The results presented above are made possible by over three decades of
work by the team of scientists and engineers who created a microcalorimeter
array for X-rays and overcame enormous setbacks. We gratefully
acknowledge the entire \xrism\ team's effort to build, launch, 
calibrate, and operate this observatory. We thank the referee for useful
comments. 
Part of this work was supported by the U.S.\ Department of Energy by Lawrence Livermore National Laboratory under Contract DE-AC52-07NA27344, and by 
NASA under contracts 80GSFC21M0002 and 80GSFC24M0006 and grants 80NSSC20K0733, 80NSSC18K0978, 80NSSC20K0883, 80NSSC20K0737, 80NSSC24K0678, 80NSSC18K1684, 80NSSC23K0650, and 80NNSC22K1922.
Support was provided by JSPS KAKENHI grant numbers JP23H00121, JP22H00158, JP22H01268, JP22K03624, JP23H04899, JP21K13963, JP24K00638, JP24K17105, JP21K13958, JP21H01095, JP23K20850, JP24H00253, JP21K03615, JP24K00677, JP20K14491, JP23H00151, JP19K21884, JP20H01947, JP20KK0071, JP23K20239, JP24K00672, JP24K17104, JP24K17093, JP20K04009, JP21H04493, JP20H01946, JP23K13154, JP19K14762, JP20H05857, and JP23K03459, the JSPS Core-to-Core Program, JPJSCCA20220002, and the Strategic Research Center of Saitama University.
LC acknowledges support from NSF award 2205918. 
CD acknowledges support from STFC through grant ST/T000244/1. 
LG acknowledges support from Canadian Space Agency grant 18XARMSTMA.
NO acknowledges partial support by the Organization for the Promotion of Gender Equality at Nara Women's University. 
MS acknowledges support by the RIKEN Pioneering Project Evolution of Matter in the Universe (r-EMU) and Rikkyo University Special Fund for Research (Rikkyo SFR). 
AT acnowledges support from the Kagoshima University postdoctoral research program (KU-DREAM). 
SY acknowledges support by the RIKEN SPDR Program. 
IZ acknowledges partial support from the Alfred P.\ Sloan Foundation through the Sloan Research Fellowship.
DN acknowledges funding from the Deutsche Forschungsgemeinschaft (DFG)
through an Emmy Noether Research Group (grant number NE 2441/1-1).

\end{acknowledgments}

\appendix

\section{Modeling velocity structure function and its scatter}
\label{app:vsf}

Under the assumptions of isotropy and homogeneity, all three
components of the gas velocity field ${\bf v}({\bf x})$ have the same power 
spectrum $P$\/ and $P({\bf k}) = P(k)$. We further assume 
that the velocity field can be represented by a Gaussian random field with a
power spectrum $P(k)$ given by eq.\ (\ref{eq:pk}).

What is measured by \xrism\ is not ${\bf v}({\bf x})$ but the first ($\mu_z$)
and second ($\sigma_z$) moments of the projected velocity component aligned
with the line of sight, given by (assuming 
the line of sight is along the $z$-axis in a Cartesian coordinate system):
\begin{eqnarray}
\mu_z(x, y) &=& \int{v_z({\bf x})\,\varepsilon({\bf x})}\,dz \label{eqn:proj_mu} \\
\sigma_z^2(x, y) &=& \int{v_z^2({\bf x})\,\varepsilon({\bf x})}\,dz - \mu_z^2
\label{eqn:proj_sig}
\end{eqnarray}
where 
\begin{equation}
\varepsilon({\bf x}) = \frac{{\rm EM}({\bf x})}{\displaystyle\int{{\rm EM}({\bf x})\,dz}} \label{eqn:proj_em}
\end{equation}
is the emission measure $\rm{EM}$ normalized by its LOS
integral. We have assumed that the Coma ICM is approximately
isothermal, so the velocity weighting is independent of the gas
temperature. Consequently, the 2D power spectrum of $\mu_z$ 
is related to the 3D spectrum $P$\/ as
\begin{equation}
P_{\rm 2D}(k) = \int{P\left(\sqrt{k^2+k_z^2}\right)P_\varepsilon(k_z)}\,dk_z,
\end{equation}
where $P_{\varepsilon}(k_z)$ is the 1D power spectrum of the normalized
emission measure \citep{Zhuravleva2012, ZuHone2016, Clerc2019}.

It is not possible to measure $P_{\rm 2D}$ directly from a limited set of
pointings such as ours; however, a more easily computed quantity, the 
second-order structure function VSF($r$) of the bulk velocity, is related 
to the power spectrum. As shown by \citet{ZuHone2016} and \citet{Clerc2019}:
\begin{equation}
{\rm VSF}(r) \,\equiv\, \langle|\mu_z(\boldsymbol{\chi}+{\bf r})-\mu_z(\boldsymbol{\chi})|^2\rangle \,=\,
4\pi\int_0^\infty[1-J_0(2\pi{kr})]\,P_{\rm 2D}(k)\,k\,dk,
\end{equation}
where ${\bf r}$ is the distance vector on the sky between a pair of points,
and the average is taken over all points with position
$\boldsymbol{\chi}$. $J_0$ is a Bessel function of the first kind of order
zero. It can similarly be shown that the LOS velocity dispersion
$\sigma_z$ is related to $P$\/ by
\begin{equation}
\sigma_z^2 = \int{P}(k)\,[1-P_{\varepsilon}(k_z)]\,d^3{\bf k}.
\end{equation}
In light of these relationships, we can constrain the parameters of the 
power spectrum $P(k)$ from the structure function ${\rm VSF}(r)$ calculated
from the line shift measurements,
and the measured velocity broadening $\sigma_z$. In practice, these formulas
need to be corrected for the shape of the pixelization, region shapes, and
PSF \citep{Clerc2019}, which we do by replacing $P_\varepsilon$ with
$P_{\varepsilon}P_{\rm inst}$, where $P_{\rm inst}$ is the power spectrum of
these instrumental effects. To model the PSF effect, we use a Gaussian with
$\sigma=34''$ (HPD=1.3\am), which has the same scattering fraction 
for the adjacent 1.5\am\ square regions and flat brightness
distribution as the actual \resolve\ PSF. This approximation is 
adequate given other significant uncertainties.   

The construction of ${\rm VSF}(r)$ must account for the statistical,
systematic, and cosmic variance uncertainties of the velocities. 
As noted in \S\ref{sec:vels}, the
statistical error on the bulk velocity $\mu_z$ is $\sigma_{\rm stat}$ =
30 \kms\ for the four quadrants in the central pointing and for the
entire southern pointing, and the systematic error (from gain uncertainty)
is $\sigma_{\rm sys} \lesssim 15$ \kms\ (\S\ref{sec:data}). As
shown by \citet{ZuHone2016} and \citet{Cucchetti2019}, the observed
structure function ${\rm VSF}'(r)$ is biased by these errors as:
\begin{equation}
{\rm VSF}'(r) = {\rm VSF}(r) + 2\sigma_{\rm stat}^2 + 2\sigma_{\rm
sys}^2.
\label{eqn:sf_obs}
\end{equation}
Additionally, the observed velocity field in Coma represents one possible
realization of the random field model with power spectrum given by $P(k)$. 
To determine the expected sample, or ``cosmic,'' variance on
${\rm VSF}(r)$, we adopt a procedure similar to \citet{ZuHone2016} and
construct 3D cubes of Gaussian random velocity fields using a range of
values for $P_0$, $\ell_{\rm inj}$, $\ell_{\rm dis}$, and $\alpha$. For a given
set of model parameters, we perform 1500 realizations of the 3D velocity
field and project it using eqs.\ \ref{eqn:proj_mu}-\ref{eqn:proj_em} to
obtain realizations of $\mu_z$ and $\sigma_z$. From these realizations, we
compute ${\rm VSF}'(r)$ using eq.\ \ref{eqn:sf_obs} 
and determine the asymmetric sample variance (i.e., taking the variance of
deviations above and below the mean separately) on ${\rm VSF}'(r)$.
We find that, to an adequate precision, the
combined statistical and sample variance on ${\rm VSF}'(r)$ follows a
power-law distribution, which we use to compute the total variance for any
value of ${\rm VSF}'(r)$. These total variances and the mean values of 
${\rm VSF}'(r)$ for each set of the $P(k)$ model parameters are then 
compared to the observed VSF values to calculate and minimize $\chi^2$. 
For the sample variance on $\sigma_z$, we use eq.\ 7 from \citet{Clerc2019}. 
We validated this model fitting procedure using random realizations 
of the velocity field, from which we successfully derived the input parameters.

\section{Predictions for kinetic pressure from cosmological simulations}
\label{app:cosmosim}

To compare \xrism\ observations with theoretical predictions for ICM
kinematics, we primarily use the TNG-Cluster suite of cosmological 
simulations \citep{Nelson24}, selecting Coma-like systems 
and emulating \xrism\ measurements of gas velocities for direct comparison
with observations. TNG-Cluster consists of 352 high-resolution
MHD re-simulations of clusters extracted from a 1 Gpc parent box of a dark 
matter-only simulation, including radiative cooling, magnetic fields, 
and stellar and AGN feedback processes \citep{Nelson19, Pillepich18b, Pillepich18a}.

We identify Coma-like systems among the simulated clusters at $z=0$ based 
on halo mass and core properties, selecting clusters with $M_{\rm
200c}\approx10^{15.1-15.4}M_\odot$, which corresponds to the Coma mass 
range \citep{Ho22}. From those, we select only non-cool-core systems with 
central cooling times exceeding $7.7$ Gyr \citep{Lehle24}. In
TNG-Cluster, central cooling time is strongly correlated with
cluster's dynamical state. We verify that non-cool-core clusters are the most
disturbed, typically undergoing mergers. We end up with a sample of 14 
Coma-like systems. For each, we compute the emission-weighted
velocity dispersion and temperature for the
central pointing within a 90 kpc square aperture, corresponding to
the \resolve\ FOV at the Coma redshift. We consider three projections 
along the $x$, $y$, and $z$\/ directions of the simulation domain as 
separate clusters. The LOS velocity dispersion and temperature for the 
central pointing are given by
\begin{equation}
\sigma_{\rm los}^2 = \frac{\sum_{i}\varepsilon_{i}{\rm v}_{{\rm los},i}^2}{\sum_{i}\varepsilon_{i}}-\bigg(\frac{\sum_{i}\varepsilon_{i}{\rm v}_{{\rm los},i}}{\sum_{i}\varepsilon_{i}}\bigg)^2,
\end{equation}
\label{eqn:sigma}
and
\begin{equation}
T = \frac{\sum_i \varepsilon_i T_i}{\sum_i \varepsilon_i},
\label{eqn:kT}
\end{equation}
where ${\rm v}_{{\rm los},i}$ ($T_i$) is the LOS velocity (temperature) of
the $i^{\rm th}$ gas cell, and $\varepsilon_i$ is its X-ray emissivity in
the 6--8 keV energy range. The X-ray emission is computed using the
density, temperature, and metallicity of each simulated gas cell, employing 
the {\sc apec} plasma model \citep{Smith01}. The computed $\sigma_{\rm los}$ 
and $T$\/ are used to calculate the kinetic-to-total pressure ratio
$p_{\rm kin}/p_{\rm tot}$ as described in eq.\ (\ref{eq:pkin}).

In addition to the TNG-Cluster predictions, we compare the \xrism\ results
with previous numerical studies of the ICM pressure ratio $p_{\rm
kin}/p_{\rm tot}$ in cluster cores \citep{Lau09, Nelson14, Vazza18,
Angelinelli20, Sayers21, Groth25}, as shown in Fig.\ \ref{fig:pturb}. 
These predictions are derived from cosmological simulations of massive 
clusters that utilize various numerical schemes and physical models for 
the ICM. Unlike our TNG sample, clusters in those 
studies were not selected to match both of our Coma-like criteria. 
For simulations presented in \cite{Lau09, Sayers21, Groth25}, we can 
extract predictions specific to non-relaxed clusters, whereas the other 
studies do not distinguish clusters based on their dynamical state. 
Additionally, the pressure ratios $p_{\rm kin}/p_{\rm tot}$ are not computed
using emission-weighted averaging --- most present mass-weighted averages, 
while \cite{Sayers21} derives the pressure ratio from a three-dimensional 
triaxial analysis of the cluster mass distribution. Despite these caveats, 
these studies predict core pressure ratios in the 5--10\% range, 
consistent with the TNG-Cluster predictions, albeit systematically
toward the higher end of the TNG-Cluster range.

{}

\allauthors
\end{document}